\documentclass[twocolumn,preprintnumbers,amsmath,amssymb,prl ]{revtex4-1}

\usepackage{times}

\usepackage{amsmath}
\usepackage{amsfonts}
\usepackage{amssymb}
\usepackage{graphicx}
\usepackage{color}

\definecolor{mypink}{RGB}{219, 48, 122}
\definecolor{mygreen}{RGB}{51, 153, 102}
\definecolor{brown}{RGB}{165, 42, 42}

 \newcommand{\kx}{$k_\mathrm{x}$}
 \newcommand{\ky}{$k_\mathrm{y}$}
\newcommand{\kxy}{$k_\mathrm{x,y}$ }

\newcommand{\kz}{$k_\mathrm{z}$}
\newcommand{\hn}{$\mathrm{h\nu}$}

 \newcommand{\kpar}{$k_\parallel$\ }

\newcommand{\GMo}{$\overline{\Gamma\mathrm{M}}$}

 \newcommand{\GKo}{$\overline{\Gamma\mathrm{K}}$}

 \newcommand{\MGM}{$\overline{\mathrm{M}\Gamma\mathrm{M}}$}
 
 \newcommand{\GZ}{$\Gamma\mathrm{Z}$}

\newcommand{\ZU}{$\mathrm{Z}$-$\mathrm{U}$}

\newcommand{\ZAG}{$\mathrm{Z}$-$\mathrm{A}$-$\Gamma$}
 
\newcommand{\UZU}{$\mathrm{U'}$-$\mathrm{Z}$-$\mathrm{U}$}

\newcommand{\aGeTe}{$\alpha$\nh GeTe}
\newcommand{\bGeTe}{$\beta$\nh GeTe}
\newcommand{\aaGeTe}{$\alpha$\nh GeTe(111)}

\newcommand{\TPsm}{$\mathrm{TP}^{\sigma}_-$}
\newcommand{\TPsp}{$\mathrm{TP}^{\sigma}_+$}
\newcommand{\TPp}{$\mathrm{TP}^{+}_-$}
\newcommand{\TPm}{$\mathrm{TP}^{-}_-$}

\newcommand{\n}[1]{$n$\nobreakdash-\hspace{0pt}}
\newcommand\nh{\mbox{-}}

\newcommand\dg{$^{\circ}$}

\newcommand{\Pz}{$P_{z}$}
\newcommand{\Pxy}{$P_{x,y}$}
\newcommand{\Px}{$P_{x}$}
\newcommand{\Py}{$P_{y}$}
\renewcommand{\lll}{$\langle 111 \rangle$}

\begin{document} 

\title{Triple point fermions in ferroelectric GeTe}

\author{Juraj Krempask{\'y}$^{1}$}
\author{Laurent Nicola\"i$^{2}$}
\author{Martin Gmitra$^3$}
\author{Houke Chen$^{4}$} 
\author{Mauro~Fanciulli$^{5}$}
\author{Eduardo~B.~Guedes$^{1}$}
\author{Marco~Caputo$^{1}$}
\author{Milan~Radovi{\'c}$^{1}$}
\author{V.~V.~Volobuev$^{6,7}$}
\author{Ond\v rej Caha$^{8}$}
\author{Gunther Springholz$^{9}$}
\author{Jan Min{\'a}r$^{2}$}
\author{J.~Hugo~Dil$^{1,10}$}

\affiliation{
$^{1}$Photon Science Division, Paul Scherrer Institut, CH-5232 Villigen, Switzerland\\
$^{3}$New Technologies-Research Center University of West Bohemia, Plze{\v n}, Czech Republic\\
$^{3}$Institute of Physics, P. J. {\v S}af\'arik University in Ko{\v s}ice, Park Angelinum 9, 040 01 Ko{\v s}ice, Slovakia\\
$^{4}$Department of Physics, Tsinghua University, Beijing 100084, China\\
$^{5}$LPMS, Universit\'e de Cergy-Pontoise, 95031 Cergy-Pontoise, France\\
$^{6}$International Research Centre MagTop, Institute of Physics,
Polish Academy of Sciences, Aleja Lotnikow 32/46, PL-02668 Warsaw, Poland\\
$^{7}$National Technical University "KhPI", Kyrpychova Str. 2, 61002 Kharkiv, Ukraine\\
$^{8}$Department of Condensed Matter Physics, Masaryk University, Kotl\'a\v rsk\'a 267/2, 61137 Brno, Czech Republic\\
$^{9}$Institut f{\"u}r Halbleiter-und Festk{\"o}rperphysik, Johannes Kepler Universit{\"a}t, A-4040 Linz, Austria\\
$^{10}$Institut de Physique, \'{E}cole Polytechnique F\'{e}d\'{e}rale de Lausanne, CH-1015 Lausanne, Switzerland\\ 
}

\date{\today}

\begin{abstract}
Ferroelectric \aGeTe\ is unveiled to exhibit an intriguing multiple non-trivial topology of the electronic band structure due to the existence of triple-point and type-II Weyl fermions, which goes well beyond the giant Rashba spin splitting controlled by external fields as previously reported. Using spin- and angle-resolved photoemission spectroscopy combined with ab initio density functional theory, the unique spin texture around the triple point caused by the crossing of one spin degenerate and two spin-split bands along the ferroelectric crystal axis is derived. This consistently reveals spin winding numbers that are coupled with time reversal symmetry and Lorentz invariance, which are found to be equal for both triple-point pairs in the Brillouin zone. The rich manifold of effects opens up promising perspectives for studying non-trivial phenomena and multi-component fermions in condensed matter systems. 
\end{abstract}

\maketitle

The exploration of topological materials commenced with the discovery of topological insulators \cite{Kane_Mele_graphene_2005b, TI_1,TI_2,TI_3,TI_4,TI_5,TI_6,TI_7, Hasan_Kane_RMP_2010, Bradlyn_Science_2016}, followed by topological semimetals \cite{HDing_TaAs_PRX,Soluyanov_2015,Armitage_RMP_2018,Hasan_Science_2015}, topological superconductors \cite{Hao_FeSe_2018} and nodal line semimetals \cite{Fang_2016}. Recently, topological semimetals with \hbox{triply-degenerate} points were predicted in materials such as molybdenum phosphide \cite{HDing_MoP}, tungsten carbide \cite{HDing_WC_2018}, InAs$_{1-x}$Sb$_x$ alloys \cite{Winkler_PRL_2016}, and strained HgTe \cite{Zaheer_PRB_2013}. In contrast to an even number of band crossings, band structures with triple band crossing are rather rare in condensed matter. Such triple-point (TP) fermions can be found along the high symmetry axis in the Brillouin zone (BZ) allowing for two and one dimensional double group representations \cite{Winkler_PRL_2016,Winkler_ChPB_2019, Zhu_PRX_2016}. This is also the case for \aaGeTe\ which was so far extensively studied in terms of the topologically trivial Rashba-type band structure \cite{Picozzi_AdvM,JK_PRB, JK_JPChS:2017}. 

The origin of these intriguing features in \aGeTe\ lies in the spontaneous structural phase transition of the inversion symmetric rock-salt \bGeTe\ structure (FCC, space group \#225) into a ferroelectric rhombohedral structure with Ge-atoms displaced by $\approx$0.3 $\mathrm{\AA}$ and distorted by $\approx$2\dg\ (space group \#160) \cite{JK_PRB, JK_MDPI_2019}. This inversion symmetry breaking introduces a spin-splitting except for the bands that line up with the high-symmetry ferroelectric \lll\ axis, which is the main symmetry axis of the $C_{3v}$ point group. Along this axis TP fermions can form which are classified as \hbox{type-A} or B, depending on the number of nodal lines that connect a pair of TP.

\begin{figure}[ht!]
\includegraphics[width=8.6 cm]{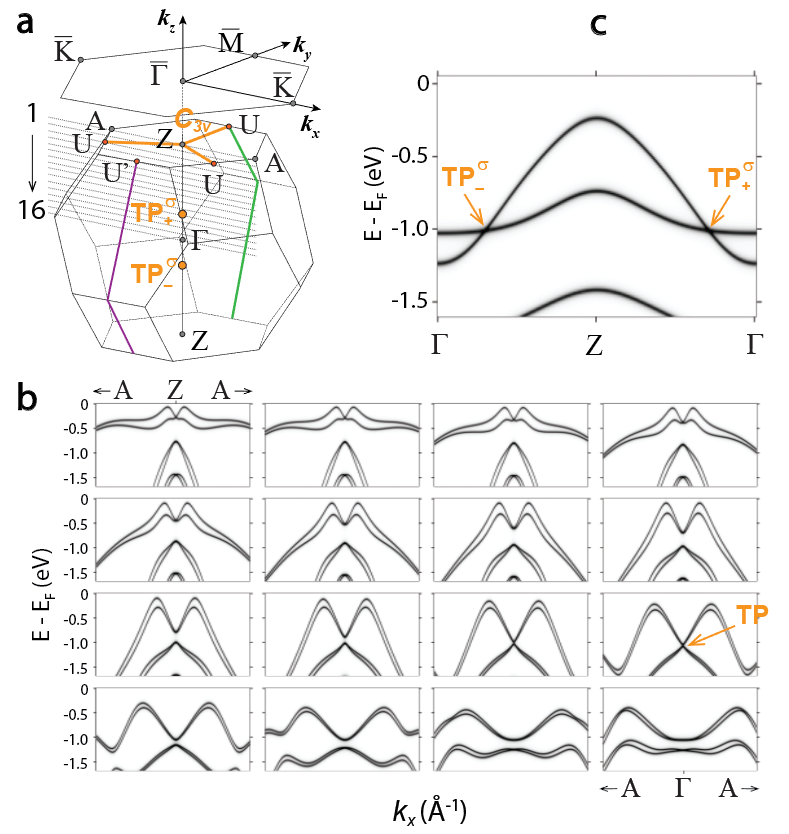}
\caption{(a) Bulk BZ and surface BZ of rhombohedral \aGeTe\ with \lll\ ferroelectric axis along the \kz\ direction. (b) BSF initial state calculations along \kx\ for sixteen selected \kz\ values denoted in panel (a). (c) Band structure along \kz\ with position of \TPsm\ and \TPsp\ indicated by orange arrows.
}   
\label{Fig1}
\end{figure}

In this Letter we provide theoretical and experimental evidence that \aGeTe\ possess type-B triple point fermions inside its valence band states along the ferroelectric \lll\ axis. To this date such non-trivial topology in \aGeTe\ failed to be noticed. The spin texture around the TP is unravelled and the same spin winding number of $\pm 1$ is found for both TP pairs in the 3D BZ. Furthermore, we identified additional Dirac and \hbox{type-II} Weyl fermions that coexist with these three-component fermions. 

Epitaxial films of 200~nm of \aGeTe\ for our (spin-) and angle-resolved photoemission spectroscopy ((S)ARPES) studies were grown by molecular beam epitaxy on BaF$_\mathrm{2}$(111) substrates and in situ transferred for high-resolution ARPES at the UE112-PGM2a beamline of BESSY II. SARPES measurements were performed at the Swiss Light Source of the Paul Scherrer Institut at the COPHEE end station equipped with two orthogonal Mott polarimeters to measure the three spatial components of spinors in arbitrary reciprocal space points \cite{Hoesch_JESRP}. We compared our experimental data with calculated initial state band-maps presented in terms of Bloch spectral functions (BSF) as implemented in the fully relativistic spin-polarised Korringa-Kohn-Rostoker theory \cite{Minar_RPP} and with the full potential all-electron Wien2K density functional code. Considering the relatively small energy scales involved, this allowed us to check the results for consistency.  

\begin{figure}[hb!]
\includegraphics[width=8.7 cm]{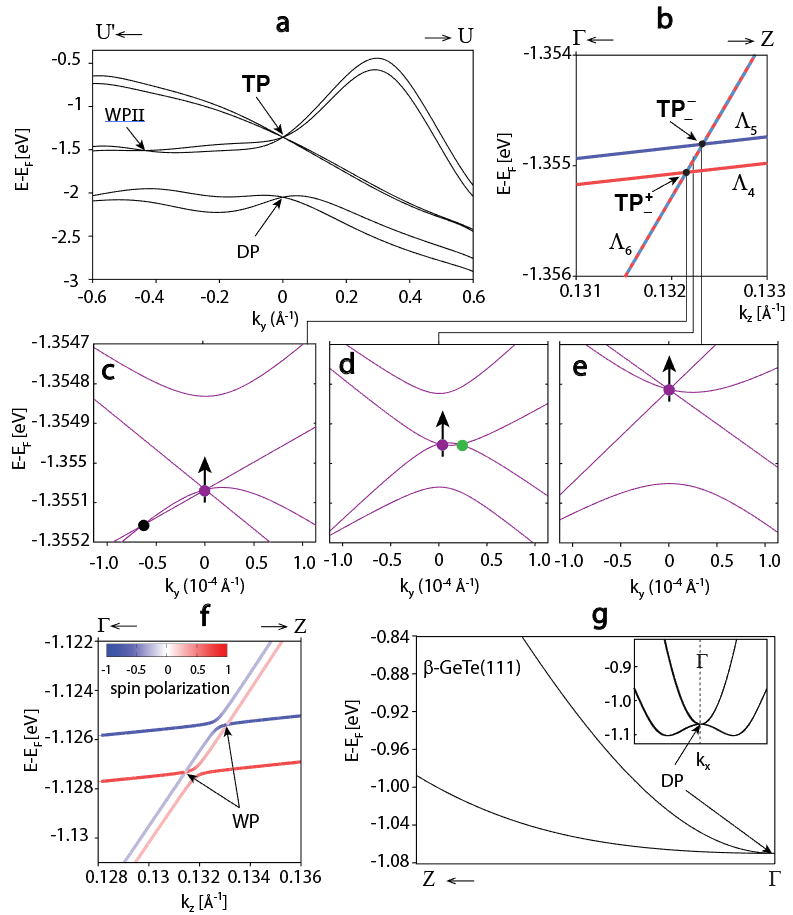}
\caption{(a) Calculated band structure along the \UZU\ mirror plane, with triple point (TP), type-II Weyl fermion (WP II) and Dirac point (DP). (b) Formation of TP pair (\TPm, \TPp) by spin-degenerate band $\Lambda_6$ and two spin split bands $\Lambda_{4,5}$. (c,d,e) Close-up of band structure along \ky\ at three different values of \kz\ across the TP pair. Three types of touching points are indicated with black, green, and magenta markers, black arrows indicate the \kz\ dispersion of the latter. (f) Band structure around the TP under 10 T B-field along the z-direction. The TP splits into a pair of Weyl points (WP). (g) Band structure for rock-salt \bGeTe\ above ferroelectric phase transition. The triple fermions merges into a single Dirac point (DP). (inset) band structure along \kx.
}   
\label{Fig2}
\end{figure}

Fig.~\ref{Fig1}a shows the BZ with two pairs of triple points $\mathrm{TP}^{\sigma}_z$, where the $\sigma=\pm$ index denotes the TP within a single pair, and $z=\pm$ denotes the two TP pairs with opposite $k_z$ within the BZ. Panel (b) shows valence band (VB) dispersions for sixteen selected $k_z$ values in the \ZAG\ plane. The top-left image presents the Rashba-type spin splitting of the uppermost VB at the Z-point. When moving $k_z$ towards $\Gamma$ these bands disperse and finally meet with the lower spin-split bands at approximately $\frac{1}{3}$BZ distance from the $\Gamma$-point (orange arrow). Another visualization of the TP formation is in panel (c) plotting the dispersion along \GZ.  At first glance this crossing point is reminiscent of the topologically protected touching point between electron and hole pockets in a double-Weyl scenario\cite{Takane_2019,Sanchez_2019,Rao_2019}, but Fig.\,\ref{Fig2}b-e unambiguously resolve a three-component fermion matching the space group symmetry.
Besides this, the bulk electronic structure in the \UZU\ mirror plane for a $k_z$ right at the TP in Fig.\,\ref{Fig2}a shows two additional band crossings with non-trivial topology. First, there is a Dirac point (DP) near 2 eV binding energy. Second, a \hbox{type-II} Weyl point (WP II) appears around -0.4 \AA$^{-1}$ as indicated in the figure. The presence of this Weyl point influences the spin texture and changes the spin winding number from $\pm$1 to $\pm$2, as seen in Fig.\,\ref{Fig5}e,f and discussed in more detail below.

\begin{figure}[ht!]
\includegraphics[width=8.8cm]{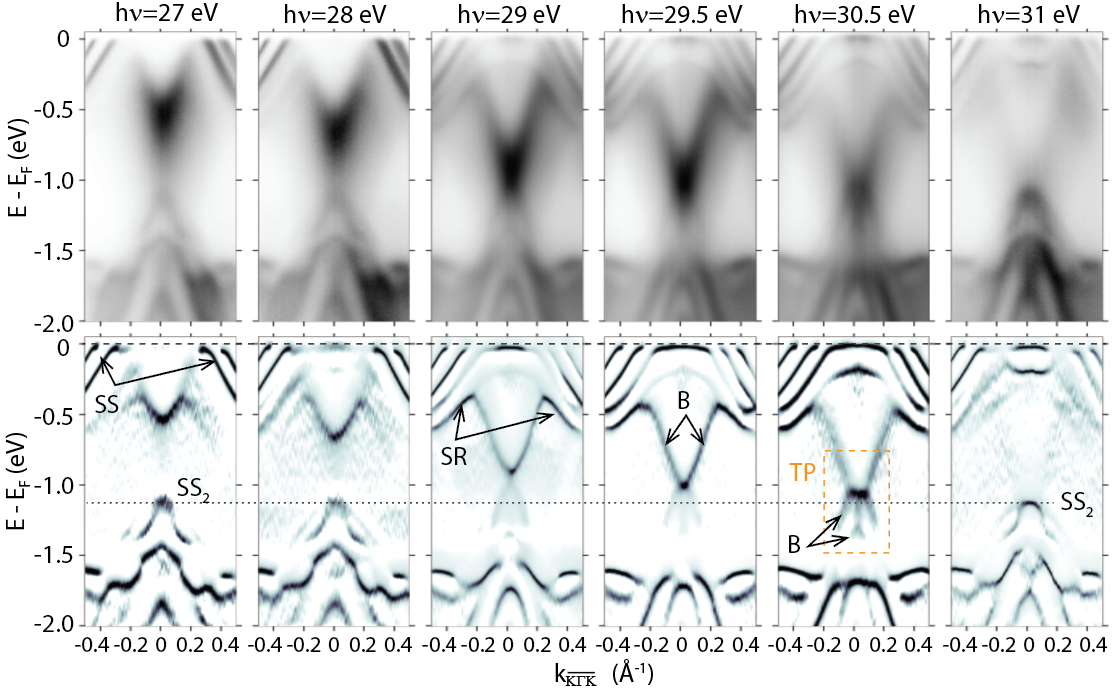}
\caption{
(top) High-resolution ARPES data for various photon energies around the TP. (bottom) Corresponding curvature maps showing the TP formation at 30.5 eV photon energy inside the orange frame. The arrows indicate surface states SS, $\mathrm{SS_2}$, surface resonances (SR) and bulk bands (B).}   
\label{Fig3}
\end{figure}

Following the TP-classification by Winkler $et~al.$\,\cite{Winkler_ChPB_2019}, \aGeTe\ should host a \hbox{type-B} TP with four nodal lines. Fig.~\ref{Fig2} summarizes this scenario: panel (b) resolves the bands $\Lambda_{4,5,6}$ dispersing in \kz, whereby the two spin-split bands $\Lambda_4$, $\Lambda_5$ cross the doubly degenerate $\Lambda_6$, forming a pair of triple fermion quasiparticles. In contrast to \hbox{type-A} TPs with only one nodal line, the triple points within each pair $\mathrm{TP}^{\sigma}_-$ and $\mathrm{TP}^{\sigma}_+$ are connected by four nodal lines. One nodal line is lined up with the $C_{3v}$ axis and therefore shows up at zero in-plane momentum only (purple markers in Fig.~\ref{Fig2}c-e). On the other hand the three remaining nodal lines are extending in three vertical mirror planes, which in our case line up along the three \ZU\ directions (orange lines in Fig.\,\ref{Fig1}a). Panels (c-e) are close-ups of the bands structure across the single TP$^{\pm}_{-}$ pair for three $k_z$ values. For the intermediate $k_z$ between \TPm\ and \TPp\ there are two crossing points forming two of the four nodal lines, whereby the one at finite \kpar\ (black and green markers) disperse with \kz\ (see animated Figure S1 in Supplemental Material).

As a whole, the triple fermion forms an intermediate state between Dirac and Weyl fermions. Actually, as indicated in Fig.\,\ref{Fig2}f, \aGeTe\ inside a 10\,T magnetic field (which is many orders of magnitude lower than the exchange field in (GeMn)Te \cite{JK_GMT}) the TP splits into a pair of Weyl points. Furthermore, above the ferroelectric transition temperature ($\mathrm{T_C}$\,$\approx$700\,K) the TP transforms into a Dirac point as seen in panel (g). This renders GeTe rather unique as all three phases can be achieved under realistic conditions in a single material system.

In our earlier \aGeTe\ ARPES studies we extensively described the measured electronic structure in terms of surface states (SS), surface resonances (SR), and bulk states (B) \cite{JK_PRB,JK_GMT,JK_PRB,JK_JPChS:2017,Kremer_PRR_2020,JK_PRR_2020}. Such a decomposition also applies to the ARPES data in Fig.\,\ref{Fig3} and allows to find the appropriate region for measuring the TP and its spin texture. Below we demonstrate that inside the $\mathrm{-0.2<k_{y}<0.2\,\AA^{-1}}$ momentum span the additional spin tag in SARPES data enables us to unambiguously resolve the individual bulk bands around a TP-pair.

Based on the \kz-series of band maps in Fig.\,\ref{Fig1}b, the ARPES data in Fig.\,\ref{Fig3}c, and the schematic \kz\ vs. photon energy diagram (Supplemental Figure S2), the TP is predicted, and experimentally verified, for photon energies around $\mathrm{h\nu}$ = 30, 58 and 89\,eV. The latter two values correspond to consecutive TP pairs at opposite parts of the BZ, thus enabling us to experimentally examine their symmetry connection.

\begin{figure}[b!]
\includegraphics[width=8.8 cm]{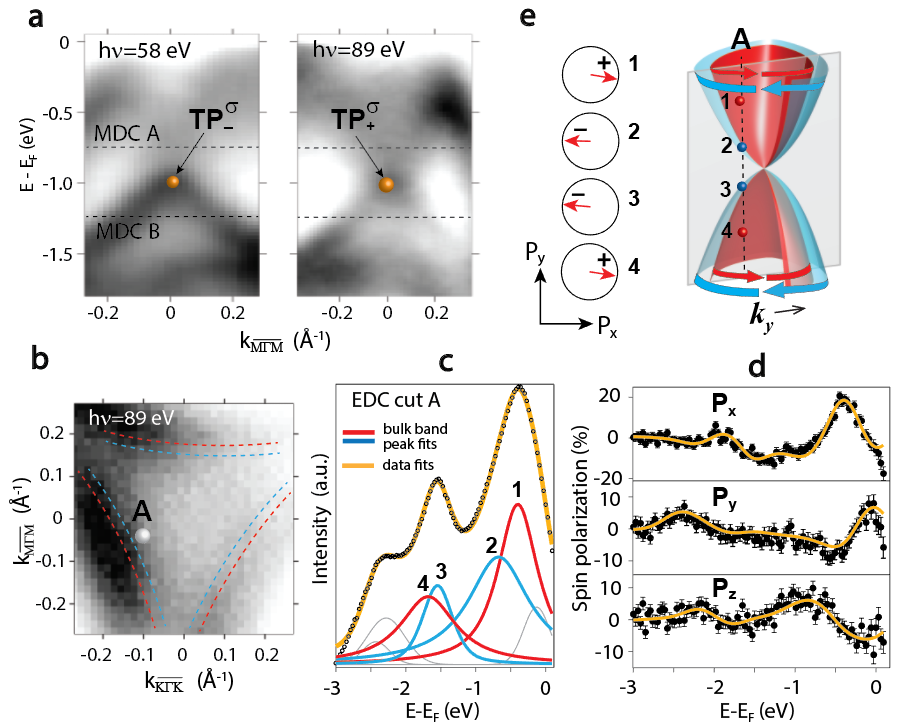}
\caption{
(a) ARPES band maps along \MGM\ measured at the COPHEE setup with \hn=58 and 89\,eV. Orange bullets indicate \TPsm\ and \TPsp. MDC\,A and MDC\,B indicate where data in Fig.\,\ref{Fig5} is obtained. (b) CEM measured at \hn=89 eV and -0.4 eV binding energy at COPHEE. Marker A indicates the \kxy\ locus used for SARPES EDC, next to the bulk bands denoted with dashed red-blue lines. (c,d) SARPES EDC cut\,A total intensity and related $P_{x,y,z}$ spin polarizations with corresponding $\mathrm{B_{1,2,3,4}}$ bulk band fitting indicated in red-blue and surface-derived bands in gray. (e) 3D TP cartoon illustrating considered EDC cut A with corresponding \hbox{in-plane} \hbox{spin-textures}, showing spin vectors projected mainly along \Px\ with a characteristic (+\,-\,-\,+) pattern around the TP.}   
\label{Fig4}
\end{figure}

\begin{figure*}[ht!]
\includegraphics[width=\textwidth]{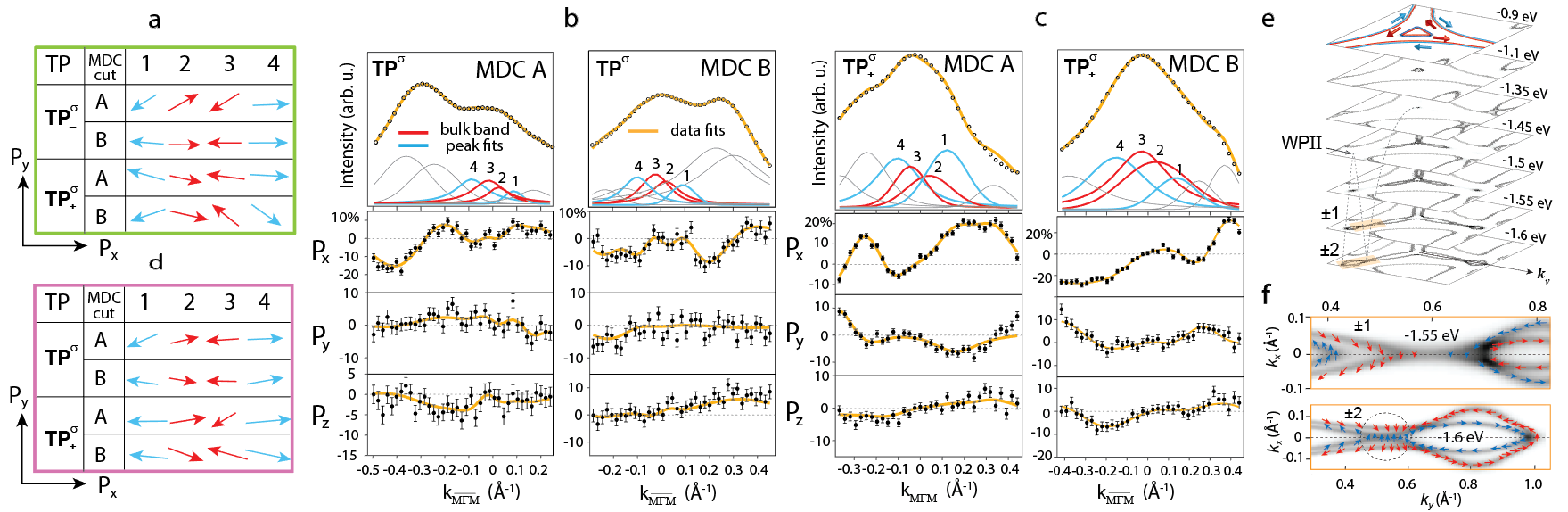}
\caption{(a) Tabulated spin textures for \TPsm\ and \TPsp\ measured along the MDC\,A (MDC\,B) cuts 250 meV above (below) the TP, as denoted in Fig. 4a. (b,c) Corresponding SARPES intensities and 3D spin polarization fitting for \TPsm\ and \TPsp\, respectively. Red and blue curves indicate the $\mathrm{B_{1,2,3,4}}$ peak fitting, surface-derived bands are in gray. (d) Analogous spin textures measured by rotating the sample by 180\dg, the frames in green and magenta indicate two inequivalent directions denoted in Fig.\,\ref{Fig1}a. (e) BSF constant energy maps, black arrow indicates the \hbox{Weyl-II} fermion, dashed line its energy dispersion and hybridization with TP bands. (f) Close ups of spin-resolved CEMs at -1.55 and -1.6\,eV, respectively, showing a change in spin winding number from $\pm$1 to $\pm$2 around the \hbox{WP II} point inclusion with conspicuous "spin swirl" denoted by dashed circle.}   
\label{Fig5}
\end{figure*}

A notable feature of the TP electronic structure is that the bands gradually shift between \TPm\ and \TPp\ as marked by black arrows in Fig.~\ref{Fig2}c-e. This spectral weight transfer across the TP is manifested in a series of high-resolution ARPES band maps in Fig.\,\ref{Fig3}c, and their curvature plots \cite{Zhang:2011curv} in the bottom panels. In accord with the theoretical predictions (Fig.~\ref{Fig1}b), the \kz-dispersion of the spectral weight is progressing toward the TP around 30.5\,eV, where it becomes equally distributed above and below the TP. Interestingly, the band maps inside the orange frame for \hn=30.5\,eV appear to mimic the TP band mixing seen in Fig.~\ref{Fig2}c-e, however with much larger momentum and energy scale.
Because the bulk spectral intensity around the TP totally suppresses the weakly contributing surface states $\mathrm{SS_2}$ (horizontal dashed line in Fig.\,\ref{Fig3}), the contribution of the latter can be neglected in our SARPES analysis. We note that surface states are typically associated with nodal points in topological metals and semimetals, but $\mathrm{SS_2}$ is not interconnecting the TP pairs because it is clearly off-set in energy at zero momentum, as indicated inside the orange frame. In fact, because both TP pairs line up in the \GZ-direction, the associated surface states cannot show up on the (111) surface.   

Similar to the Rashba model, the spin polarization of bulk states around a TP is expected to be tangentially momentum-locked in accord with time-reversal symmetry (TRS) properties, as also shown in the calculated spin texture in Fig.\ref{Fig5}e. Our SARPES data is visualized as spin-resolved energy distribution (EDC) and momentum distribution curves (MDC) summarized in Figures\,\ref{Fig4} and \ref{Fig5}, respectively. For all measurements our analysis concentrates on the in-plane spin polarisation \Pxy\ of the four spin-split bulk bands denoted by $B_{1,2,3,4}$. Their properties are determined using a 3D vectorial fit of the measured spin expectation values with a well established fitting routine that extracts their positions, widths, intensities, and 3D spin vector \cite{Fabian_NJP}. For the EDCs a Shirley background is subtracted from the total intensity, whereas for the MDCs a constant background is used. To simplify comparison between theory and experiment, we aligned the sample in such a way that the out-of-plane spin polarization \Pz\ is oriented along the \kz-direction and \Px\ (\Py) along the \GKo\ (\GMo) direction. Within this experimental geometry a tangential spin winding is determined by \Px.

We first test the spin texture in terms of TRS and Lorentz invariance (LI) for the selected \kxy point labeled A in Fig.\,\ref{Fig4}b, with SARPES data summarized in panels (c-d). The resulting spin texture is summarized in panel (e) next to a  model visualizing the EDC cut through the four $B_{1-4}$ bulk bands around the TP. We note that the spin helicity dictated by TRS and LI are preserved: the spins are nearly tangentially locked in momentum, whereby bands $B_{1,4}$ and $B_{2,3}$ do not change the spin helicity. The spin texture imposed by LI, i.e. symmetric spin in energy with regard to the TP, becomes apparent when considering the \hbox{(+\ -\ -\ +)} pattern in Fig.\,\ref{Fig4}d; in contrast to the Rashba-type \hbox{(+\ -\ +\ -)} pattern seen in Supplemental Figure S3, measured at the BZ boundary Z-point (see Fig.\,\ref{Fig1}a).

This experimental result is further confirmed in Fig.\,\ref{Fig5}. We measured each TP-pair with two MDC cuts 250\,meV above (A) and below (B) the TP, as depicted in Fig.\,\ref{Fig4}a. In addition, because there are two inequivalent directions (Fig.\,\ref{Fig1}a green and magenta lines), each TP-pair was measured along U'-A-U and after rotation by 180\dg along U-A-U' (see SARPES data in Supplemental Figure S4). The resulting in-plane spin textures shown in Fig.\,\ref{Fig5}d are consistent with a spin winding number of $\pm 1$. To best of our knowledge this is the first experimental evidence of the spin texture of a triple fermion, featuring a winding number of $\pm$1 in accord with both TRS and LI and theoretical predictions. Here it should be noted that the inclusion of WP II induces an additional spin swirl and changes the winding number to $\pm 2$ as illustrated by the calculations in Fig.\ \ref{Fig5}e. Furthermore, the two triple point pairs TP$^\sigma_-$ and TP$^\sigma_+$ show the same spin helicity in accordance with the absence of a $(x,y)$ mirror plane.

In conclusion, we demonstrated that \aGeTe\ is a type-B triple fermion topological system enabled by the ferroelectric non-centrosymmetric atomic arrangement of Ge and Te atoms along the \lll\ axis. Furthermore, we measured for the first time the spin winding number for a triple point system. Moreover, \aGeTe\ is currently unique in the possibility to manipulate topological properties by temperature variation \cite{JK_PRB}, ferroelectric switching \cite{JK_PRX}, and/or magnetic doping \cite{JK_GMT}. 

M.F. acknowledges support from the SNF Project No. P2ELP2\_181877 and G.S. from the Project No. P30960-N27 of the Austrian Science Fund (FWF). J.M. and L.N. would like to thank the CEDAMNF Project financed by the Ministry of Education, Youth and Sports of Czech Republic, Project No. CZ.02.1.01\/0.0\/0.0\/15\_003/0000358 and the Czech Science Foundation (GACR), Project No. 2018725S. M.G. acknowledges VEGA 1\/0105\/20 and Internal Research Grant System VVGS-2019-1227, V.V.V. the Foundation for Polish Science through the IRA Programme co-financed by EU within SG OP, and finally O.C. acknowledges the financial support by the European Regional Development Fund Project CEITEC Nano+ (No. CZ.021.01\/0.0\/0.0\/16\_013\/0001728).


%

\end{document}